%% file: dmann.tex
\documentclass[10pt,sigplan,letterpaper,twocolumn]{article}
\usepackage[10pt,nocopyright]{sigmin}

\usepackage{listings}
\usepackage{color}
\usepackage[hyphens]{url}                                         
\usepackage[breaklinks,colorlinks]{hyperref}                      
\usepackage[usenames,dvipsnames]{xcolor}                          
\hypersetup{citecolor=blue,linkcolor=blue}                        
\usepackage{amsmath,amsopn,amssymb,amsthm}                        
\usepackage{thmtools}
\usepackage{thmbox}
\usepackage[toc,page]{appendix}
\usepackage{subcaption}                                           
\usepackage{endnotes,microtype,xspace,fancyvrb,multirow} 
\usepackage{epigraph} 
\usepackage{diagbox} 

\usepackage{booktabs}      
\usepackage{tabularx}
\usepackage{array,underscore,relsize}                             
\usepackage[T1]{fontenc}                                          
\usepackage{times}                                                
\usepackage{fancyhdr,lastpage}                                    
\usepackage{enumitem}                                             
\usepackage[labelfont=bf,font=normal,skip=1pt]{caption}           
\usepackage[export]{adjustbox}                                    
\usepackage{breakurl}                                             
\usepackage{pifont}                                               
\usepackage{tablefootnote}                                        
\usepackage{bbding}                                        
\usepackage[linesnumbered,vlined,boxed,commentsnumbered]{algorithm2e}
\DontPrintSemicolon
\newcommand{\ra}[1]{\renewcommand{\arraystretch}{#1}}

\usepackage{algpseudocode}
\usepackage{amssymb}
\usepackage[normalem]{ulem}
\usepackage[hang,flushmargin]{footmisc}
\usepackage{textcomp}
\usepackage{graphicx}
\usepackage{authblk}

\usepackage[compact]{titlesec}
\titleformat*{\section}{\large\bfseries}
\titleformat*{\subsection}{\normalsize\bfseries}
\titleformat*{\subsubsection}{\normalsize\bfseries}
\titlespacing{\section}{0pt}{3ex}{1ex}
\titlespacing{\subsection}{0pt}{2ex}{1ex}

\hypersetup{
  colorlinks,
  linkcolor={red!50!black},
  citecolor={blue!50!black},
  urlcolor={blue!80!black}
}

\newcommand{\name}{\textsc{pvdb}}

\newcommand{\Vdb}{VectorDB}
\newcommand{\vdb}{vectorDB}

\newcommand{\topk}{\textit{top-k}}
\newcommand{\topc}{\textit{top-c}}

\newcommand{\fig}[1]{Figure{~\ref{#1}}}
\newcommand{\eg}{e.g.,~}
\newcommand{\ie}{i.e.,~}

\newcommand{\one}{\texttt{\uppercase\expandafter{\romannumeral1}}}
\newcommand{\two}{\texttt{\uppercase\expandafter{\romannumeral2}}}

\newcommand{\red}[1]{\textcolor{red}{#1}}

\newcommand{\SJ}[1]{\textcolor{blue}{SSJ: #1}}

\newcommand{\TODO}[1]{\textcolor{red}{TODO: #1}}

\newcommand{\stitle}[1]{\vspace{1.1ex}\noindent{\bf #1}}

\usepackage{tikz}

\usepackage{balance}
\begin{document}
\RestyleAlgo{ruled}


\title{\Large \bf{
  Characterizing the Dilemma of Performance and Index Size in Billion-Scale Vector Search and Breaking It with Second-Tier Memory
}}

\setlength{\affilsep}{0.5em}

\author[1,2]{Rongxin Cheng}
\author[1]{Yifan Peng}
\author[1,2]{Xingda Wei\thanks{Xingda Wei is the corresponding author (\url{wxdwfc@sjtu.edu.cn})}}
\author[1]{Hongrui Xie}
\author[1,2]{Rong Chen}
\author[3]{Sijie Shen}
\author[1] {Haibo Chen}
\affil[1]{\vspace{-2.mm}Institute of Parallel and Distributed Systems, SEIEE, Shanghai Jiao Tong University\vspace{0.8mm}}
\affil[2]{Shanghai AI Laboratory\vspace{-1.mm}}
\affil[3]{Alibaba Group\vspace{-1.mm}}

\date{}
\maketitle

\frenchspacing

\input{abs}
\input{body/intro}
\input{body/bg-v1}
\input{body/analysis-v2}
\input{body/random}

\input{body/graph}

\input{body/cluster}

\input{body/overall}

\input{body/related}

\input{concl}

\balance

\small{
\bibliographystyle{acm}
\bibliography{dmann}
}

\clearpage

\end{document}

%% file: abs.tex
\begin{abstract}

\noindent
Vector searches on large-scale datasets 
are critical to modern online services like web search and RAG,
which necessity storing the datasets and their index on the secondary storage like SSD. 
In this paper, we are the first to characterize the 
trade-off of performance and index size in existing SSD-based graph and cluster indexes: 
to improve throughput by {5.7\,$\times$} and {1.7\,$\times$}, these indexes have to pay a {5.8\,$\times$} 
storage amplification and {7.7\,$\times$} with respect to the dataset size, respectively.
The root cause is that
the coarse-grained access of SSD mismatches 
the fine-grained random read required by vector indexes with small amplification.

This paper argues that second-tier memory, such as remote DRAM/NVM connected via RDMA or CXL, 
is a powerful storage for addressing the problem from a system's perspective,
thanks to its fine-grained access granularity.
However, putting existing indexes---primarily designed for SSD---directly on second-tier memory cannot fully utilize its power.
Meanwhile, second-tier memory still behaves more like storage, 
so using it as DRAM is also inefficient.
To this end, we build a graph and cluster index that centers around the performance features of second-tier memory. 
With careful execution engine and index layout designs, we show that vector indexes 
can achieve optimal performance with orders of magnitude smaller index amplification,
on a variety of second-tier memory devices.

Based on our improved graph and vector indexes on second-tier memory,
we further conduct a systematic study between them to facilitate developers choosing the right index for their workloads.
Interestingly, the findings on the second-tier memory contradict the ones on SSDs.

\noindent

\end{abstract}

%% file: body/intro.tex
\section{Introduction}
\label{sec:intro}

\noindent
Multi-dimensional vectors with tens or hundreds of dimensions
are powerful representations 
of multi-modal data including but not limited to text, vision, and videos~\cite{DBLP:journals/corr/abs-1301-3781,DBLP:conf/emnlp/PenningtonSM14}.
As a result, vector search---searching a database with numerous vectors 
to find the closest one given a query vector---is a key pillar 
in supporting real-world tasks 
like web search~\cite{DBLP:conf/kdd/LiLJLYZWM21} and retrieval augmented generation (RAG)~\cite{DBLP:conf/nips/LewisPPPKGKLYR020}.
However, finding the exact closest vector is often impractical, 
particularly for high-dimensional and large-scale vector datasets 
commonly found in real workloads~\cite{DBLP:conf/sosp/XuLLXCZLYYYCY23,DBLP:conf/compgeom/Clarkson94}. 
Therefore, systems typically use \emph{approximate nearest neighbor search} (ANNS) 
to find an approximation of the $k$ closest vectors in the dataset (i.e., find the {\topk})~\cite{DiskANN,vbase,SPANN,SPTAG,DBLP:conf/sosp/XuLLXCZLYYYCY23}. 

A key system requirement for vector search is high performance.
For example, vector search is the backend of Google search~\cite{google-vector-search},
a service that is latency sensitive~\cite{DBLP:journals/micro/BarrosoDH03,DBLP:journals/cacm/DeanB13} and demands high processing rate: 
e.g., handles 8.5 billion queries each day~\cite{google-query}.
To accelerate ANNS-based vector search, people have built two types of indexes:
graph~\cite{DiskANN,hnsw,nsg} and cluster~\cite{SPANN, faiss, DBLP:conf/sosp/XuLLXCZLYYYCY23}.
Graph indexes use a configurable number of edges to link vectors that are close in distance,
 and conduct graph traversal to locate the {\topk} vectors. 
Cluster indexes group closed vectors into clusters. 
They then perform a brute-force search on a configurable number of clusters to find the {\topk} of a query vector.

For vector searches on large-scale datasets,
it is common to store the graph and clusters on secondary storage like SSD~\cite{SPANN,DBLP:conf/sosp/XuLLXCZLYYYCY23,DiskANN}.
As the SSD bandwidth is several orders of magnitude slower than DRAM, 
SSD-based vector search is \emph{I/O bound}. 
For example, a commodify {64}-core server can search {3,624--5,212\,Mvectors/sec} 
with SIMD\footnote{\footnotesize{Single Instruction Multiple Data.}} on common vector datasets with 100--384 dimensions. 
In comparison, an SSD with {5.3\,GB/s} bandwidth can only search {14--53\,Mvectors/sec} 
even considering its bandwidth is fully utilized (e.g., in cluster indexes). 
For indexes like graph, fully utilizing the SSD bandwidth is impossible 
due to the small random reads (e.g., 256\,B).

\stitle{The high performance and low index size dilemma for I/O-bounded vector searches.}
One intuitive way to improve I/O performance in existing vector indexes 
is to increase the index size, 
e.g., store extra information (edges and replicated vectors).
For the graph, when the size of the graph is increased by adding more edges, 
the number of hops---each corresponding to a small random I/O---dramatically decreases. 
Meanwhile, as each hop reads more data (e.g., from 256 to 1024\,B), 
the SSD bandwidth utilization also increases.
For the cluster,
by replicating vectors in adjacent clusters, 
the number of clusters required for the search~\cite{SPANN,DBLP:conf/sosp/XuLLXCZLYYYCY23} also dramatically reduced,
so the overall search performance improves. 

However, the index size required for optimal vector search is orders of magnitude 
higher than the traditional scalar indexes. 
For example, DiskANN~\cite{DiskANN} and SPFresh~\cite{DBLP:conf/sosp/XuLLXCZLYYYCY23}---the state-of-the-art graph and cluster indexes---require 
indexes that are  
{4.9}\,$\times$ and {6.7}\,$\times$ larger than the indexed dataset (index amplification)
 for an optimal performance, respectively (\textsection{\ref{sec:motiv}}). 
In comparison, 
scalar indexes like B$^+$Tree or hash table only have an 
 index amplification of {0.4--1.2}\,$\times$~\cite{DBLP:conf/sigmod/ZhangAPKMS16}\footnote{\footnotesize{The database community still believes that such an amplification is huge~\cite{DBLP:conf/sigmod/ZhangAPKMS16}.}}.
 Unlike scalar indexes, the index size, 
 the number of I/Os and the size of each I/O are determined by the machine learning 
    algorithms that build the index and search it,
which we argue can hardly improve to match the requirements of high-performance SSD accesses.

Another way to tackle this problem is from a system's perspective without changing the algorithms.
For example, we can deduplicate the replicated vectors in cluster indexes 
by replicating the address instead of vectors (\textsection{\ref{sec:design-cluster-basic}}), 
thereby reducing the cluster index with replication by {84\%}. 
However, naively changing the index layout could result in 
orders of magnitude more small random reads (100--384\,B) to read replicated vectors, 
and the resulting performance is only {1.5\%} of the original replicated design on SSD.

\stitle{Root cause: workloads mismatch SSD requirements.}
The workloads of graph or vector indexes with small amplification share the same characteristics: 
they are I/O intensive, require many random accesses over large datasets, 
and the sizes of reads are small, 
e.g., a few edges in the graph (e.g., 256\,B) and a single replicated vector (e.g., 100--384\,B) for cluster deduplication. 
The above workloads mismatch with the high bandwidth utilization requirements of 
traditional storage like SSD, i.e.,
using a few sufficient large reads (4\,KB) to fully utilize the SSD bandwidth.

\stitle{The second-tier memory for the rescue. }
We argue that 
second-tier memory---volatile or non-volatile (NVM) memory 
that is attached to the host with fast interconnects like RDMA and CXL~\cite{cxl} (\textsection{\ref{sec:random}}),
opens up opportunities to systematically address the problem.
Specifically, 
these devices behave like storage but support finer access granularity (256\,B vs. 4\,KB), 
which matches the workload patterns of vector indexes. 
Moreover, they are robust to random reads with even smaller access granularity (e.g., 100\,B).
Specifically, adding {50\%} such I/O might not affect the memory bandwidth utilization, 
while in SSD the utilization drops by up to {43\%}.
\emph{
Therefore, we can trade a few sequential accesses in vector search
for dramatically reducing index size. 
}

\stitle{Challenges and solutions for utilizing second-tier memory.}
In this paper, 
we built two indexes around the performance features of second-tier memory
to show the effectiveness of vector indexes in second-tier memory.
On various devices including RDMA, CXL, and NVM, 
our graph index can reach the optimal performance with {4--44\%} less index storage, 
and our cluster index can achieve so with {40\%} index size amplification.

Achieving so is non-trivial.
First, existing billion-scale vector indexes are designed for SSDs,
so they cannot fully utilize the fine-grained access nature of second-tier memory.
Second, treating the second-tier memory as DRAM also results in poor performance 
because they still behave like storage devices,
e.g., with an order of magnitude higher latency and should minimize the number of small reads.
For example, placing the graph index on second-tier memory would shift the performance bottleneck from I/O to computation, 
but the relatively long access latency of second-tier memory hinders the CPU from fully utilizing the SIMD for vector search (\textsection{\ref{sec:graph}}).
Therefore, we retrofit the execution pipeline of the graph index 
to hide the execution delay caused by reading the second-tier memory.
Meanwhile, naively deduplicating the cluster index would result in numerous small random I/Os on the second-tier memory.
This prevents us from fully utilizing the I/O bandwidth (\textsection{\ref{sec:cluster}}). 
To this end, we designed a grouping mechanism to minimize the small random I/Os on second-tier memory.

\stitle{An end-to-end study on comparing graph and vector indexes (\textsection{\ref{sec:study}})}.
To conclude our study of utilizing second-tier memory for vector indexes,
we conducted a systematic study comparing the graph and cluster indexes on second-tier memory.
We draw several interesting findings that contradict common findings on SSD.
First, it is widely known that graph is slower than the cluster because it is unfriendly to the SSD's I/O, 
which causes its poor performance~\cite{SPANN,DBLP:conf/sosp/XuLLXCZLYYYCY23}.
However, graph can have much better performance than the cluster,
not only because its I/O is more efficient on second-tier memory, 
but also it reads far fewer vectors than the cluster.
Second, the cluster index is notoriously for huge index size ({6.3--7.7}\,$\times$ of original dataset) caused by replication~\cite{DBLP:journals/corr/abs-2401-02116}. 
On the second-tier memory, with our improved design, 
the index size can be kept consistently small ({1.1--1.4}\,$\times$).

\stitle{Contributions. } In summary, our contributions are:  \\[-18pt]
\begin{itemize}[leftmargin=*,leftmargin=10pt,itemindent=0pt]
    \item We are the first to characterize the high performance and low index size amplification dilemma 
    in SSD-based indexes for large-scale vector datasets. \\[-18pt]
    \item We provide the first guideline on how to utilize second-tier memory for breaking the aforementioned problem. \\[-18pt]
    \item We built the first graph and cluster indexes on second-tier memory that achieve orders of magnitude higher performance 
    as well as orders of magnitude smaller index size. \\[-18pt]
    \item We present the first end-to-end study comparing the graph and vector indexes on second-tier memory. \\[-18pt]
\end{itemize}

%% file: body/bg-v1.tex
\section{Vector Search and Second-tier Memory}
\label{sec:bg}

\subsection{ANNS-based vector search and its indexes}
\label{sec:bg-anns}

\vspace{-1.ex}
\stitle{ANNS-based vector search.}
Given a query vector and a vector dataset, 
vector search finds the most similar vector in the dataset. 
Formally, given a vector dataset $[x_1, x_2, \ldots, x_n]$ and a query vector $x$,
where $x \in \mathbb{R}^d$,
the search will find $x_i = argmin_{x_i}\text{dist}(x, x_i)$,
where $\text{dist}$ is the distance metric, \eg{Euclidean distance}.
Due to the curse of dimensionality~\cite{DBLP:conf/compgeom/Clarkson94},
it is impractical to find the exact $x_i$ especially 
for high-dimensional vectors
commonly found in real-world applications~\cite{SPANN, DBLP:conf/sosp/XuLLXCZLYYYCY23,vbase}.
Therefore, systems conduct \emph{approximate nearest neighbor search (ANNS)} for the vector search,
which finds $K$ approximate nearest candidates for the $x_i$ (\topk{}).    
To accelerate ANNS, people have built two types of indexes\footnote{\footnotesize{To the best of our knowledge, 
no other types of billion-scale vector indices exist.    }}:    

\begin{figure}[!t]
        \centering 
        \includegraphics[scale=1.15]{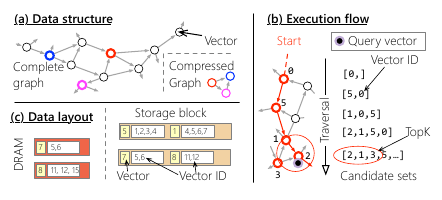}  \\[1pt]
        \begin{minipage}{1\linewidth}
        \caption{\small{%
        An illustration of 
        (a) the basic data structures of graph-based vector index, 
        (b) the execution flow of vector search with it and 
        (c) how the data structures are stored.
        }}
        \label{fig:graph}
        \end{minipage} \\[-5pt]
    \end{figure} 

\stitle{Graph-based index~\cite{DiskANN,hnsw,nsg}. }    
It stores vectors in a directed graph ({\fig{fig:graph}} (a)), where the nodes are vectors 
and the edges connect vectors that are close in distance.
For example, if $a \rightarrow b$, it means that vector $b$ is the {\topk} vector of $a$.
{\fig{fig:graph}} (b) presents the execution flow of the search. 
The search begins at a start node in the graph and identifies the nearest neighbors through 
best-first graph traversals.
Specifically, each hop in the traverse will read a vector and its edges from the storage.
The starting node of the traverse can be fixed~\cite{nsg}, chosen randomly~\cite{DPG,KGraph} 
or selected by traversing a (relative) small or compressed graph 
that is distilled from the original graph~\cite{hnsw,DiskANN,DBLP:conf/nips/0015ZL20} (see (a)).
During the traversal, the search maintains a candidate set containing the closest vectors traversed so far 
(sorted by their distances). 
After the traversal is completed, the search returns the 
{\topk} results from the candidate set as the final {\topk} results of the search.
The traversal stops as long as the candidate set size exceeds a pre-defined threshold (a hyperparameter).

The full graph is typically stored as adjacent lists in the secondary storage~\cite{DiskANN,DBLP:conf/nips/0015ZL20}, 
as shown in {\fig{fig:graph}} (c). 
Note that the nodes and edges are stored together since each traversal requires both data.
On the other hand, the compressed graph, which aids in traversal, is stored in the DRAM~\cite{DiskANN}.

The benefit of a graph index lies 
in its ability to capture the relationships between vectors 
in a fine-grained way. 
As a result, the search process has \emph{low read amplification}, 
meaning it reads only a few extra vectors beyond the required {\topk}.
The downside is that graph traversal is not friendly to storage due to 
its pointer chasing access pattern, 
resulting in long latency and low bandwidth utilization~\cite{SPANN,DBLP:conf/sosp/XuLLXCZLYYYCY23}.
Moreover, the graph index is not friendly to vector insertions, 
as building the graph necessitates reconstructing the graph~\cite{DBLP:conf/sosp/XuLLXCZLYYYCY23}.

\begin{figure}[!t]
        \centering 
        \includegraphics[scale=1.15]{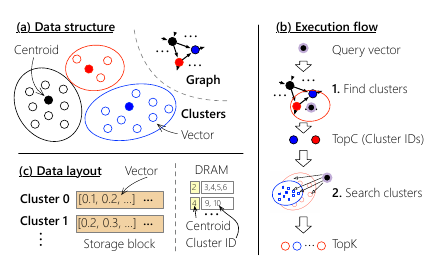}  \\[1pt]
        \begin{minipage}{1\linewidth}
        \caption{\small{%
        An illustration of 
        (a) the basic data structures of cluster-based vector index, 
        (b) the execution flow of vector search with it and 
        (c) how the data structures are stored.
        }}
        \label{fig:ivf}
        \end{minipage} \\[-5pt]
    \end{figure}

\stitle{Cluster-based index~\cite{SPANN, faiss, DBLP:conf/sosp/XuLLXCZLYYYCY23}. }    
The cluster index has been optimized for storage access. 
As shown in {\fig{fig:ivf}}, 
The vectors are partitioned into clusters 
where each cluster can be stored efficiently in a storage block (e.g., an SSD block), 
as shown in {\fig{fig:ivf} (c)}.
The search then only needs to read the closest clusters to get the {\topk}. 
To facilitate finding the closest clusters, 
cluster indexes typically use a graph index 
to record the cluster relationships, 
where the graph node is the centroids of the clusters~\cite{SPANN,DBLP:conf/sosp/XuLLXCZLYYYCY23,ParlayANN}.
Since the graph is typically much smaller than the cluster (only contains centroids), 
it is stored in memory for efficient search.
{\fig{fig:ivf}} (b) presents the concrete execution flow. 
The search first traverses the graph to find the {\topc} clusters. 
Afterward, it reads all the cluster data and searches the {\topk} among them.

The benefits of cluster-based indexes include: 
(1) they are efficient for secondary storage as they allow large bulk reads of clusters, 
and (2) they are friendly to insertion workloads, 
as vectors can be inserted into the closest clusters 
without the need for rebuilding the whole index like graph~\cite{DBLP:conf/sosp/XuLLXCZLYYYCY23}.
However, cluster-based indexes have \emph{high read amplification}
 due to redundant vectors read in each cluster. 
Meanwhile, they also have \emph{high space amplification} 
due to replications required for high accuracy (see \textsection{\ref{sec:bg-index-build}} for more details).

\subsection{Building vector indexes}
\label{sec:bg-index-build}

\noindent
The space consumed by vector indexes is closely tied to how the indexes are constructed.
Both types of indexes are built using traditional unsupervised machine learning methods.

\stitle{Graph-based index.}
To build a graph index on a dataset of vectors, 
the builder first uses k-Nearest Neighbor (\emph{kNN}) to find the closest neighbors for each vector. 
The $k$ is a static user-configured parameter for index construction. 
Note that this $k$ is unrelated to the {\topk} required by the query.
It then connects each vector with its $k$ closest neighbors to form the graph.
Graph indexes typically chooses a small $k$ (e.g., 32~\cite{DiskANN}) 
as it could significantly amplify the space of the index (see \textsection{\ref{sec:motiv}}).
Based on the graph formed by \emph{kNN}, 
existing indexes also prune the edges using methods like 
sparse neighborhood graph (SNG)~\cite{SNG}
for improved traversal quality.

\stitle{Cluster-based index.}
First, the builder partitions the vectors into clusters using \emph{kMeans}. 
The number of clusters is selected to be orders of magnitude smaller than the number of vectors.
One issue with \emph{kMeans} is that it can cause unbalanced partitioning, 
i.e., one cluster may contain significantly more vectors than others.
Unbalanced clusters can lead to high variance in search time,
so state-of-the-art indexes~\cite{SPANN,DBLP:conf/sosp/XuLLXCZLYYYCY23} will further balance the cluster size 
using a multi-constraint balanced clustering algorithm~\cite{DBLP:conf/bigdataconf/LiuHCL0Z18}.
Another issue with \emph{kMeans} is the boundary issue. 
Specifically, boundary vectors of a cluster can be assigned to multiple neighboring clusters. 
However, \emph{kMeans} assigns them to only one cluster.
Ignoring this issue leads to searching through more clusters for accuracy, 
which further results in significant performance degradation.
Therefore, existing indexes replicate boundary vectors to a set of close clusters, 
where the number of replications is statically configured before the index build.  

\subsection{Second-tier memory}
\label{sec:bg-tiered-memory}

\begin{figure}[!t]
    \hspace{-2mm}
    \centering 
    \includegraphics[scale=1.55]{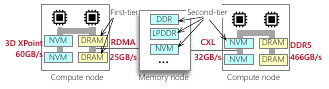}  \\[-1pt]
    \begin{minipage}{1\linewidth}
    \caption{\small{%
    First-tier memory vs. second-tier memory. 
    }} 
    \label{fig:second-tier}
    \end{minipage} \\[-5pt]
\end{figure}

\input{./body/memory-tab}

\noindent
{\fig{fig:second-tier}} presents the second-tier memory architecture that we target in this paper.
Unlike traditional fast DRAM that is directly connected to the CPU via the memory bus, 
second-tier memory is a relatively slower, but cheaper memory device 
that can be indirectly connected to the CPU through a high-bandwidth interconnect
like RDMA~\cite{racehashing,DBLP:conf/osdi/GaoNKCH0RS16} or CXL~\cite{cxl,DBLP:conf/asplos/MarufWDWABPCKC23}.
The memory can be either old-generation DRAM (DDR3), 
which is sufficient to saturate the interconnect bandwidth, 
or persistent memory like NVM~\cite{DBLP:journals/corr/abs-1903-05714,rdpma}\footnote{\footnotesize{NVM can also be directly attached to the CPU like DRAM.}}.
They have less cost per GB compared to locally attached DRAM because: 
(1) older-generation memory is cheaper
and (2) indirection allows improved memory utilization by pooling unused memory~\cite{DBLP:conf/asplos/LiBHEZNSRLAHFB23,DBLP:journals/pvldb/ZhangRLYCLWFWHB21,DBLP:conf/nsdi/GuLZCS17}.
However, they are still significantly slower than DRAM and behave more like storage (detailed in \textsection{\ref{sec:random}}):
They require a sufficient large request payload (e.g., 256\,B) to saturate the interconnect bandwidth (see \fig{fig:random-access}).

To show that our results generalize to a variety of second-tier memory, 
we considered all three known implementations in production (see Table~\ref{tab:memory}): 
RDMA-attached DRAM (RDMA), CXL-attached DRAM (CXL), and NVM. 

%% file: body/memory-tab.tex
\begin{table}[!t]
    \vspace{2.2mm}
    \begin{minipage}{1\linewidth}
        \caption{\small{
            Performance features of different memory technologies.
        }}
    \label{tab:memory}
    \end{minipage} \\[-1pt]
    \centering
    \small{
    \resizebox{.99\linewidth}{!}{
    \ra{1.25}

\begin{tabular}{l|rrrrr}
    \hline
    & \textbf{SSD} & \textbf{DRAM} & \textbf{RDMA} & \textbf{CXL} & \textbf{NVM} \\
    \hline
    \textbf{Bandwidth} & 5.3\,GB/s  & 35\,GB/s    & 25\,GB/s    & 32\,GB/s     & 60\,GB/s \\
    \textbf{Latency}   & 75\,$\mu$s & 0.1\,$\mu$s & 2.8\,$\mu$s & 0.3\,$\mu$s  & 0.4\,$\mu$s \\
    \hline
  \end{tabular}    

    }
    } 
    \end{table}  

%% file: body/analysis-v2.tex
\section{The Dilemma of High Performance and Small Index Size in Vector Indexes}
\label{sec:motiv}

\setlength{\abovedisplayskip}{2pt}
\setlength{\belowdisplayskip}{5pt}

\begin{figure*}[!ht]
        \hspace{-5pt}
        \includegraphics[left, scale=1.14]{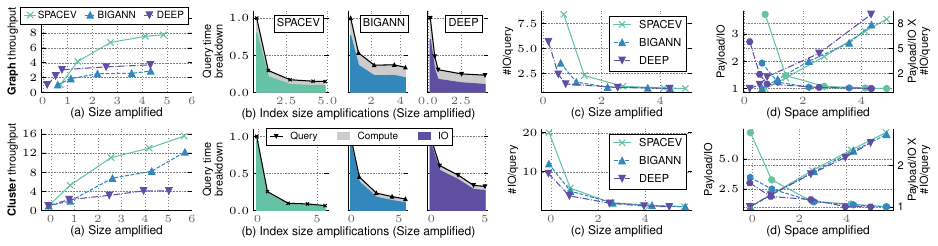} \\[5pt]
        \begin{minipage}{1\linewidth}
            \caption{\small{
                (a) Performance—index size trade-off in existing graph-based (upper) and cluster-based (bottom) vector indexes.
                The amplification is normalized to the total dataset size without indexing.                
                (b) I/O dominates index performance across various datasets.                
                (c) The performance increases when the index size becomes larger 
                is due to the reduced number of I/O sent per-query 
                and (d) the request payload per I/O and its trend compared to the reduced number of I/O per query.
                Note that all the performance data is normalized.
            }}
        \label{fig:analysis}
    \end{minipage} \\[-5pt]
\end{figure*}

\input{body/dataset-tab}

\vspace{-1.ex}
\stitle{Setup, datasets and baseline indexes.} 
We conducted experiments on a machine equipped with an Intel D7-P5520 SSD (7.7\,TB, 5.3\,GB/s), 
and an Intel Xeon Gold 6430 CPU (20 cores, 3.4\,GHz).
We use three popular billion-scale workloads used by 
existing works~\cite{DiskANN,SPANN,DBLP:conf/sosp/XuLLXCZLYYYCY23}---the largest 
publicly available datasets for vector search (see Table~\ref{tab:dataset}).

We choose two vector search systems as our baselines: \\[-15pt]
\begin{itemize}[leftmargin=*,leftmargin=10pt,itemindent=0pt]
    \item \textbf{Graph (DiskANN~\cite{DiskANN})}: 
    DiskANN is the state-of-the-art graph index. 
    It follows the graph index design described in \textsection{\ref{sec:bg-anns}},
    and further leverages product quantization (PQ) to compress the graph 
    in memory to accelerate graph search. \\[-15pt]

    \item \textbf{Cluster (SPFresh~\cite{DBLP:conf/sosp/XuLLXCZLYYYCY23})}: 
    SPFresh is state-of-the-art cluster index that utilizes: 
    (1) an in-memory graph index on centroids to accelerate finding search clusters 
    for the query vector, and 
    (2) a balanced partitioning algorithm to ensure low variances in cluster sizes. 
    Additionally, SPFresh supports in-place updates.    \\[-15pt]
\end{itemize}

\noindent
Both indexes are optimized with SSD access, e.g., they use SPDK~\cite{spdk} to 
achieve the best I/O performance on SSDs,
and leverage SIMD for execution. 

\stitle{Search accuracy and runtime index hyper parameters. }
The accuracy of the vector index is defined in terms of \emph{recall}, 
i.e., the ratio of result vectors returned by the index in the ground-truth {\topk}.
Given the same accuracy, the performance of vector search is strongly 
influenced by the configuration of its runtime hyperparameters, i.e.,
the number of clusters to search and the maximum size of the candidate set 
(that determines the graph traversal hops,see {\fig{fig:graph}} (b)).

Like prior work~\cite{SPANN,DBLP:conf/sosp/XuLLXCZLYYYCY23,DiskANN}, 
we choose the parameters that achieve the best performance at a given accuracy level, 
and use an overall accuracy of at least 90\% across all experiments.
Note that for a given dataset and index, its hyperparameters are configured only once 
and remain unchanged during its query time. 

\subsection{Case study: graph index}
\label{sec:motiv-graph}

\begin{figure}[!t]
    \hspace{-0mm}
    \includegraphics[scale=0.8]{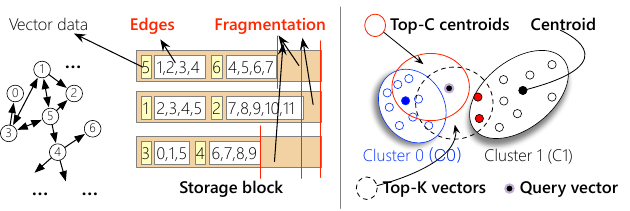} \\[5pt]
    \begin{minipage}{1\linewidth}
            \caption{\small{
            An illustration of (a) the index size amplification problem 
            in graph indexing and
             (b) why a cluster index requires replication.
            }}
    \label{fig:space-analysis}
\end{minipage} \\[-5pt]
\end{figure} 

\vspace{-1.ex}
\stitle{Sources contribute to the index size. }
Similar to a traditional graph, 
the additional space occupied by the graph vector index comes from two sources:
the space used to store graph edges, 
and the internal fragmentation that prevents cross-storage block graph traversal (see {\fig{fig:space-analysis} (a)}). 
The space for edges is non-trivial and dominates the amplification: 
For every vector in the dataset, 
the graph stores edges as adjacency list next to it (the graph is sparse). 
For instance, DiskANN by default assigns 32 edges to each vector.
Given that each edge is a 4\,B vector ID, 
the size of the list can exceed the original data (e.g., 128 vs. 100 in SPACEV). 
Specifically, DiskANN by default introduces an {0.5--1.4\,$\times$} space amplification in existing datasets. 
Note that this relatively small amplification trades graph traversal performance, 
which we will elaborate on later. 

Unlike other graph systems like graph databases and analytics, 
where the number of edges cannot be controlled by the system,
the graph vector index can configure different numbers of edges per vector, 
providing control over the space used by the graph.
Specifically, we can adjust the $k$ in the KNN algorithm used to build the graph (see \textsection{\ref{sec:bg-index-build}}) 
to control the number of edges per vector.

\stitle{Characterizing the problem. }
Ideally, we should minimize the space amplification caused by edges.
Unfortunately, the fewer the edges, the longer the path a query vector must traverse to find its approximate {\topk}.
Since an SSD access is orders of magnitude slower than computing the distance between two vectors (\eg{70\,$\mu$s vs. 1\,ns} on our platform), 
such extra hops significantly slow down the search in graph index, see {\fig{fig:analysis}} (b).
Therefore, the graph index faces a fundamental dilemma of high performance and low index size.
More specifically, since I/O dominates the graph search, 
the performance of the graph index can be characterized as:  \\[1pt]
\begin{equation}
    \label{eq:graph}
    \text{Throughput}_{search} = \text{Band.}\, /\, (\text{\#Hops} \times \text{Size}_{node+edge})
\end{equation}  

\noindent
For our evaluated dataset, the payload read ($\text{Size}_{node+edge}$) per graph traversal (Hop) 
cannot exceed the SSD access granularity (4\,KB) even if we increase the number of edges,
because otherwise would result in a large index size that is $39\,\times$ larger than the original dataset,
e.g., using 100\,B for the node and 3,996\,B to store the edges. 
As a result, 
the achieved bandwidth (Band.) divided by ($\text{Size}_{node+edge}$)  
can be simplified as the IOPS of the storage device---the maximum requests per second 
that the SDD can handle.
So we can simplify the performance model~\ref{eq:graph} as: 
\begin{equation}
    \label{eq:graph-iops}
    \text{Throughput}_{search} = \text{IOPS}\, /\, \text{\#Hops}
\end{equation}    
\noindent
The model clearly shows that when we increase the number of edges of graph indexes, 
the number of hops required for traversal decreases,
so the throughput increases.

The upper half of {\fig{fig:analysis}} (a) 
shows the empirical vector search performance respective to the index size 
(measured with size amplification) for all datasets.  
Take SPACEV dataset as an example, other datasets share a similar result.
When the index size increases from {1.7}--{5.9}\,$\times$, 
the throughput increases from {1.9--15.2}\,Kreqs/sec, and the peak 
throughput is {1.9}$\times$ faster than the DiskANN's default setup ({\#Edges/vector=32}). 
The performance improvement is due to the reduced number of hops required for traversal (see {\fig{fig:analysis} (b)}):
the number of I/O per query decreased significantly from {350} to {37}.
Even though each I/O reads more data from the SSD,
which increases from {164--580\,B} for the index when sizes grow from {1.7--5.9}\,$\times$ of the size of SPACEV,
the performance still dramatically improves.
This is because the IOPS remains steady (see {\fig{fig:analysis}} (d)), 
since such payloads are still far from saturating the SSD bandwidth.

\subsection{Case study: cluster index}
\label{sec:motiv-cluster}

\vspace{-1.ex}
\stitle{Sources contribute to the index size. }
Unlike a graph index, cluster index should ideally have negligible index size.
This is because (1) 
the index data used for recording clusters 
are orders of magnitude smaller than the total datasets 
(one cluster for every tens or hundreds of vectors)
and 
(2) clusters stored on the SSD are more resilient to cross block accesses 
(since each cluster may cross block due to its large size)
so no padding is necessary.
Unfortunately, state-of-the-art cluster indexes~\cite{SPANN,DBLP:conf/sosp/XuLLXCZLYYYCY23} extensively 
replicate a single vector across multiple clusters to address 
the lose of accuracy caused by the \emph{boundary issue} (see {\fig{fig:space-analysis} (b)}).
In this example, the {\topk} of the query vector spans the boundaries of cluster 0 and 1. 
If the {\topc} centroids only contain cluster 0, 
then the red vectors from cluster 1 will be missed from the final {\topk}, 
causing reduced accuracy.

Replicating the boundary vectors to both clusters addresses the issue,
at the cost of additional storage used for the index.
In general, the higher the replication factor, the better the accuracy.
Unfortunately, the number of replications that can be achieved is limited 
because the replication is applied globally to all vectors in the dataset. 
For example, SPFresh~\cite{DBLP:conf/sosp/XuLLXCZLYYYCY23} set 
a maximum replication factor of 8, 
which in the worst case would amplify the space by 8$\times$\footnote{\footnotesize{Existing indexes 
prune replicated vectors to reduce index size~\cite{SPANN,DBLP:conf/sosp/XuLLXCZLYYYCY23}. 
However, pruning does not fundamentally change the amplification factor caused by replication, 
e.g., only reduces the amplification from 7 to {5.7--6.3} in the datasets.}}.

\begin{figure*}[!t]
        \centering
        \includegraphics[left, scale=1.14]{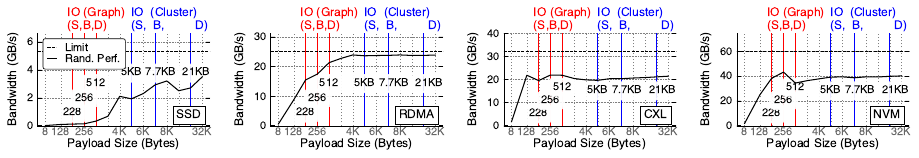} \\[3pt]
        \begin{minipage}{1\linewidth}
                \caption{\small{{
                    A comparison of peak random read performance using various read payloads 
                    for SSD and different second-tier memory hardware.  
                    Note that the x-axis is non-linear.
                    For reference, we also mark the typical read payload of the graph (red lines) 
                    and cluster (blue lines) index for SPACEV (\textbf{S}), BIGANN (\textbf{B}), 
                    and DEEP (\textbf{D}) datasets.                    
            }
            }}
        \label{fig:random-access}
    \end{minipage} \\[-5pt]
    \end{figure*} 
    
\begin{figure*}[!t]
        \centering
        \includegraphics[left, scale=1.0]{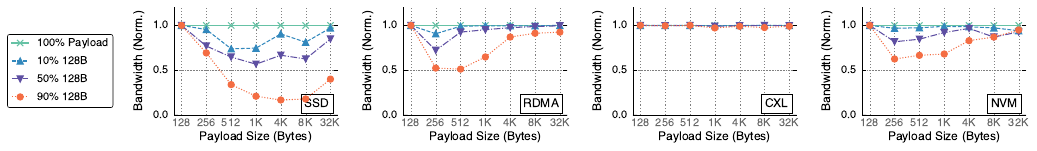} \\[-1.5pt]
        \begin{minipage}{1\linewidth}
                \caption{\small{
        An analysis of the slowdown introduced by random 128\,B reads for various devices.            
                }}
        \label{fig:hybrid-access}
    \end{minipage} \\[-5pt]
    \end{figure*}     

\stitle{Characterizing the problem. }
To avoid excessive replication, another solution 
is to increase the number of clusters searched for each query. 
For instance, if we search both cluster 0 and 1 as shown in {\fig{fig:space-analysis}} (b), 
we can achieve the same accuracy without replication.
Unfortunately, additional cluster searches obviously 
degrade the vectors searched due to the extra vectors read.

Similar to the graph index, I/O dominates the throughput of cluster vector search (the bottom of {\fig{fig:analysis}} (b)),
which can be characterized as: \\[2pt]
\begin{equation}
    \label{eq:cluster}
    \text{Throughput}_{search} = {\text{Band.}}\, /\, (\text{\#Clusters} \times \text{Size}_{cluster})
\end{equation}

\noindent
Band. is the bandwidth of the device, and {\#Clusters} is the configured number of searched clusters per query. 

Unlike the graph index, the cluster size can exceed the SSD payload threshold (4\,KB), 
so we cannot simply use the IOPS to characterize the search throughput. 
When considering the request payloads, 
the relationship between the index size and the throughput is more complex to analyze. 
As shown in Equation~\ref{eq:cluster}, increasing the index size would increase the average cluster size 
as well as reduce the number of clusters searched.
Nevertheless, empirically we found the number of clusters required to search (\text{\#Clusters}) 
decreases more than the increase in the average cluster size, 
so increasing the index size can still improve the throughput. 

The bottom of {\fig{fig:analysis}} (a) shows the performance--index size graph of cluster index. 
For BIGANN and SPACEV datasets, the throughput increase {12\,$\times$ and 16\,$\times$} 
when we increase the index size by {7.3\,$\times$ and 7.2\,$\times$}, respectively. 
For DEEP, the throughput increase {4\,$\times$} when increasing the index size by {6.7\,$\times$}.
All datasets achieve their peak throughput when configured with a replication factor of 8,
the maximum supported by SPFresh.
The performance improvement is due to fewer clusters being searched ({\fig{fig:analysis}} (c)). 
For instance, in {BIGANN}, 
it only requires {searching 20 clusters} to achieve 90\% accuracy with a replication factor of 8.
In comparison, with a replication factor 1, it has to search {240} clusters.  
Such a reduction in the number of clusters searched offsets the reduced IOPS 
caused by the increased payload per cluster (see the bottom of {\fig{fig:analysis}} (d)).

\subsection{Root cause: workloads mismatch SSD requirements}
\label{sec:motiv-root-cause}

\noindent
We attribute the dilemma to a mismatch between the 
I/O requirements for high-performance SSD access and 
I/O workloads issued by vector indexes with small index amplification. 

\stitle{Graph index requires fine-grained storage reads for practical index sizes.}
For a practical graph index with minimal index size, 
we must construct it with a small number of edges per node.
From an algorithmic perspective, 
this means that graph traversal must use random I/Os with small payloads to read these edges.
However, this fundamentally conflicts with the requirement of 
using a sufficiently large I/O payload (4\,KB) to efficiently utilize traditional 
storage devices like SSD.

\stitle{Cluster index requires irregular I/O for deduplication.}
For cluster indexes, 
we can deduplicate vectors through indirection. 
Specifically, instead of storing replicated vector data in other clusters,
 we store an address pointing to the original cluster. As the vector address (8\,B) is significantly 
 smaller than the data (128--384\,B), this approach minimizes index size.
However, it implies that each replicated vector requires a separate small random read (100--384\,B) 
to fetch the original vector.
Such an I/O pattern is irregular to the SSD: 
{50\% can cause 43\% performance drop}.

%% file: body/dataset-tab.tex
\begin{table}[t]
    \vspace{2.2mm}
    \begin{minipage}{1\linewidth}
    \caption{\small{{Billion-scale datasets used in the experiments.}}}
    \label{tab:dataset}
    \end{minipage} \\[-1pt]
    \centering
    \small{
    \resizebox{.99\linewidth}{!}{
    \ra{1.18}

\begin{tabular}{l|rrrr}
    \hline
    \textbf{Dataset} & \textbf{|Dataset|} & \textbf{Dimensions} & \textbf{Total space} & \textbf{|Queryset|} \\ \hline
    \textbf{SPACEV (S)}  &  1\,Billion                &    100 $\times$ int8              &  94\,GB               &   29\,K               \\
    \textbf{BIGANN (B)}  &  1\,Billion                &    128 $\times$ uint8             & 120\,GB               &   10\,K               \\
    \textbf{DEEP   (D)}  &  1\,Billion                &    96  $\times$ float32           & 358\,GB               &   10\,K               \\
    \hline
    \end{tabular}
    }
    }

\end{table}

%% file: body/random.tex
\section{The Power of Second-tier Memory in Vector Search}
\label{sec:random}

\noindent
We now explain our rationale for using second-tier memory for vector indexes,
whose access features perfectly match the I/O patterns of vector indexes.
\begin{enumerate}[leftmargin=*,leftmargin=10pt,itemindent=0pt]
    \item \textbf{Fine-grained block sizes\footnote{\footnotesize{We use block size to denote the smallest access granularity of second-tier memory, 
    without losing generality.}}}: 
    Compared to traditional storage like SSD, 
    second-tier memory has a more fine-grained block size (from 64--256\,B). 
    Such a pattern matches the requirements of vector indexes, 
    e.g., graph reads 228--512\,B for each traversal, 
    which means that vector index can utilize the device bandwidth more efficiently.\\[-10pt]

    \item \textbf{Robust to irregular I/Os}: 
    The measured performance is robust against 
    large payloads unaligned with the device’s access granularity,
     and random reads with not-so-small payloads (e.g., a single vector that is 128\,B),
    which is common for vector indexes with small sizes.
    Vector indexes don't issue other patterns of irregular payloads (e.g., 8\,B random reads).
\end{enumerate}

\noindent
The aforementioned advantages give us opportunities to break the space—performance dilemma in existing index designs systematically. 
First, the fine-grained I/O access granularity of second-tier memory matches the graph index requirements--—small 
random reads for low space amplification.
Second, we can utilize the robustness of irregular IO to replace (some) sequential accesses with random accesses.
Therefore, we can effectively deduplicate the cluster index storage to achieve minimal index size.

\input{./body/hardware-tab}

\stitle{Characterized second-tier memory and experiment setup.}
To show that our results generalize to various second-tier memory hardware, 
we use three machines, each with a different memory technique\footnote{\footnotesize{One machine cannot be equipped with all three techniques due to the specialized motherboard required for CXL and NVM.}}, 
 (Table~\ref{tab:hardware-info}) through all experiments. 
The machines are named after the memory technique.

We use two microbenchmarks: the first measures the random access performance of various devices, 
while the second evaluates the performance when a portion of the workload is replaced with irregular payloads, 
i.e., small 128\,B random access. 
We chose 128\,B because it represents the common payload (a single vector) used in both vector indexes (see Table~\ref{tab:dataset}).
For each device, we have carefully tuned the performance to avoid inference from underutilized hardware.
For SSD, we use SPDK to implement both benchmarks. 
For RDMA, we built index on a state-of-the-art RDMA framework~\cite{drtm-h} 
with all known RDMA-aware optimizations~\cite{DBLP:conf/nsdi/DragojevicNCH14,DBLP:journals/usenix-login/KaliaKA16,drtm-h}.
For CXL and NVM, we utilize \texttt{devdax} for direct device access 
and apply all guidelines summarized in existing studies~\cite{DBLP:conf/fast/YangKHIS20,DBLP:conf/usenix/WeiX00Z21}.

\stitle{Fine-grained access granularity that matches vector index searches.}
{\fig{fig:random-access}} shows the bandwidth achieved on different devices using various payloads.
First, SSD requires a sufficiently large payload (multiple of 4\,KB) to approach its bandwidth limit,
which is not aligned with the typical vector index payloads.
For instance, the graph index, when configured with 32 edges per node for small amplification by default,
 issues reads in 228--512\,B payloads for various datasets.
In contrast, all second-tier memory devices can saturate the bandwidth with relatively small payloads: 
RDMA, CXL, and NVM can achieve close to bandwidth limit with payloads of {4\,KB}, {128\,B}, and {256\,B}, respectively.

The differences lie in the underlying hardware design. 
Compared to RDMA, CXL, and NVM, which have block size
 of 64\,B, 64\,B and 256\,B respectively, SSD has a larger block size of 512\,B.
Additionally, issuing random access with SSD block size is insufficient to saturate the bandwidth, 
as SSDs have multiple flash dies that process 512 reads sequentially. 
Therefore, a large read payload (i.e., 4\,KB) enables the SSD to distribute reads more evenly across the dies, 
thereby improving performance with reduced die collisions~\cite{DBLP:conf/fast/JunPKKS24}. 

\begin{figure*}[!ht]
    \hspace{-5pt}
    \includegraphics[left, scale=1.1]{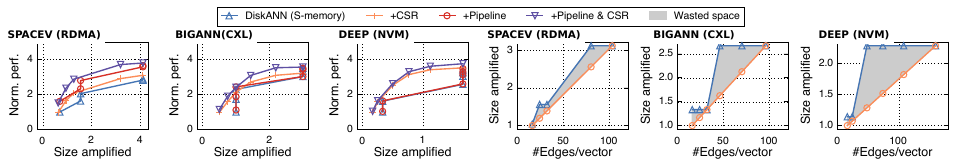} \\[4pt]
    \begin{minipage}{1\linewidth}
        \caption{\small{
        Effects of software pipeline (a)--(c) and CSR (d)--(f) on graph index. 
        Due to space limitation, we only list a subset of the overall results. 
        The trends for the others are similar. 
        }}
    \label{fig:graph-break}
    \vspace{20pt}
\end{minipage} \\[-20pt]
\end{figure*}

\stitle{Robust to irregular IO.}
SSDs are not robust to irregular payloads including:
 (1) large payloads not aligned with the device's block size (e.g., graph traversal in graph index), 
 and (2) payloads smaller than the block size (e.g., graph traversal and reading a vector). 
As shown in {\fig{fig:random-access}}, SSD performance drops when using 5\,KB, 7.7\,KB, and 21\,KB, 
which are extensively used by the graph index under different workloads.
{\fig{fig:hybrid-access}} further shows that when the ratio of 128\,B random read increases from {10--90}\% in the workloads, 
the SSD experiences {4.6--83.3\%} slowdowns.

In comparison, second-tier memory has small or no slowdown under similar scenarios.
For (1), they experience no slowdown (see {\fig{fig:random-access}}). 
For (2), as shown in {\fig{fig:hybrid-access}}, 
RDMA and NVM experience a maximum slowdown of {48.8\% and 37.5\%} under an extreme setup 
where the workloads are 50\% and 90\% workloads are 128\,B random reads.
Second-tier memory is more robust due to its fine-grained block size, 
i.e., the bandwidth wasted is minimal. 
For example, for 228\,B reads required by the graph, the worst waste is at 11\% on NVM, 
while it is 55\% on SSD, even without considering die collisions.
Finally, RDMA is less robust when handling small random access 
due to the increased network protocol overhead for each request~\cite{DBLP:conf/usenix/GoukLKJ22}. 
This suggests that reducing irregular access to second-tier memory remains necessary.

%% file: body/hardware-tab.tex
\begin{table}[!t]
    \vspace{2.2mm}
    \begin{minipage}{1\linewidth}
        \caption{\small{
        Measurement machines with second-tier memory.
        }}
    \label{tab:hardware-info}
    \end{minipage} \\[-1pt]
    \centering
    \small{
    \resizebox{.99\linewidth}{!}{
    \ra{1.3}
    \begin{tabular}{p{1.1cm}|p{4.5cm}p{3.2cm}}
        \hline
        \textbf{Platform} & \textbf{Host CPU configuration}                                      & \textbf{Second-tier Memory}       \\ \hline
        \textbf{RDMA}     & Intel(R) Xeon(R) Gold 6430 (32 cores, 3.4 GHz)     & BlueField-3 200\,Gbps (up to \textbf{25\,GB/s})            \\
        \textbf{CXL}      & Intel(R) Xeon(R) Platinum 8468V (48 cores, 3.8GHz) & MXC CXL Memory eXpander (CXL 2.0), PCIe 5.0 x8 (up to \textbf{32\,GB/s}) \\
        \textbf{NVM}      & Intel(R) Xeon(R) Gold 6330 CPU (28 cores, 2.0GHz)  & 8 $\times$ Optane 200 Series (up to \textbf{60\,GB/s})                \\
        \hline
        \end{tabular}  

    }
    } 
    \end{table}  

%% file: body/graph.tex
\section{Improved Graph Index Design}
\label{sec:graph}
\begin{figure*}[!t]
    \hspace{-5pt}
    \includegraphics[left, scale=1.1]{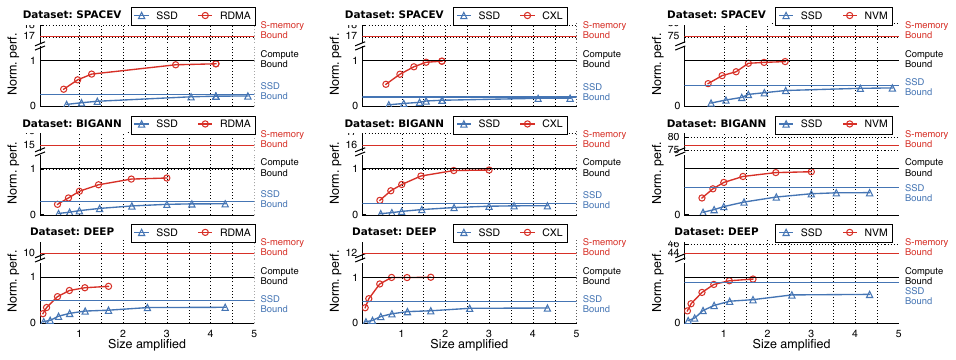} \\[-4pt]
    \begin{minipage}{1\linewidth}
        \caption{\small{
        End-to-end graph index performance. S-memory states for second-tier memory used.
        }}
    \label{fig:graph-scale}
\end{minipage} \\[-5pt]
\end{figure*}

\noindent
Putting graph indexes on the second-tier memory 
can dramatically improve the index performance thanks to its features.  
However, existing system designs for graph indexes do not match or fully utilize 
the increased I/O capability of second-tier memory,
causing performance under-utilization as well as storage waste. 
To this end, we propose two designs for graph index on the second-tier memory: \\[-15pt]
\begin{enumerate}[leftmargin=*,leftmargin=10pt,itemindent=0pt]
    \item \textbf{Software pipeline.}
    With improved I/O performance, the computing capability 
    becomes the performance bottleneck. 
    Existing synchronous search design fails to fully utilize computational power,
    so we still need a large index to amortize the second-tier memory access latency. 
    We propose a software pipeline mechanism to asynchronously process I/O with computation, 
    thereby maximizing the use of computational power. 
    \\[-15pt]

    \item \textbf{Compressed graph layout.}
    The existing graph index suffers from internal fragmentation to avoid cross-SSD block access.
    The overhead space is trivial for graphs with large amplification, 
    but it is significant when using a small-sized index on second-tier memory. 
    Observing that second-tier memory is robust against irregular I/Os with unaligned access payloads (\textsection{\ref{sec:random}}), 
    we use a compact storage layout when storing graph indexes on it. 
    \\[-15pt]
\end{enumerate}

\stitle{Software pipeline. }
The latency of second-tier memory is on the order of {300--3000\,ns}, 
which is much higher than that of first-tier DRAM ({100\,ns}).
This causes significant idle time if the CPU processes the I/O synchronously, 
as in the existing design: it busy waits for the I/O to complete before doing the computation---compare 
the distances of candidates with the vector node read back. 

To this end, we interleave memory access and vector distance computation from multiple queries 
to hide the second-tier memory access latency\footnote{\footnotesize{Interleaving does not help SSDs 
since the I/O is the bottleneck.}}.
Specifically, we schedule multiple queries execution in a per-hop way for inter-query parallelism:
after issuing a second-tier memory request, we will schedule the execution of next unfinished query, 
thereby processing different queries in a pipelined way in software. 
Note that our software pipeline is different from batching because we won’t wait for a batching window before processing a query: 
the query is processed as soon as possible. 
For RDMA, we leverage coroutines~\cite{DBLP:conf/osdi/KaliaKA16} to implement such a feature.
For CXL and NVM, since their memory accesses are implicitly issued via CPU load instructions, 
we will reorganize the instruction orders in the search 
to conduct a batch of loads before executing computation, 
as well as use \texttt{mm\_prefetch} to realize the feature.

{\fig{fig:graph-break}} shows the performance improvement of the software pipeline.
For RDMA, the software pipeline improves the performance by {1.2--1.6\,$\times$}.
On CXL and NVM, the pipeline can also improve the performance by up to 1.2\,$\times$. 
The improvements of the pipeline are more obvious in RDMA since it has the highest latency. 

\stitle{Compressed layout. }
One significant design choice in DiskANN is to use padding for storage: 
without it, DiskANN suffers a {48--58\%} performance loss on SSD.
Such a padded design might waste {4--44\%} index spaces, as shown in {\fig{fig:graph-break}} (d)--(f). 
It is unnecessary on the second-tier memory thanks to its robustness to irregular I/O that cross-block size. 
Therefore, we simply choose a Compressed Sparse Row (CSR) layout~\cite{DBLP:books/daglib/0009092} 
for storing the graph index on the second-tier memory.
As also shown in {\fig{fig:graph-break}},
CSR reduces space without compromising performance.

\begin{figure*}[!ht]
    \hspace{-5pt}
    \includegraphics[left, scale=1.1]{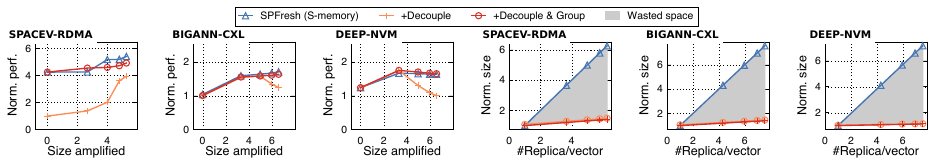} \\[3pt]
    \begin{minipage}{1\linewidth}
        \caption{\small{
        Effects of cluster-aware grouping (a)--(c) and decoupled layout (d)--(f) on graph index. 
        Due to space limitation, we only list a subset of the overall results. 
        The trends for the others are similar. 
        }}
    \label{fig:cluster-break}
    \vspace{20pt}
\end{minipage} \\[-20pt]
\end{figure*}

\stitle{Overall performance. }
{\fig{fig:graph-scale}} shows the end-to-end performance of the graph index on RDMA, CXL, and NVM
with all the above optimizations enabled\footnote{\footnotesize{Due to limited memory on our platform, 
we only report performance of 100M datasets for CXL and DEEP100M for NVM.}}.
We calculate the computation bound by performing all calculations when storing the graph in DRAM
and bandwidth bound by with the peek bandwidth measured in \fig{fig:random-access}.
On second-tier memory, we can see that fine-grained random reads won't cause significant bandwidth waste,
so graph index can reach the CPU's computation bound with a small index size amplification 
({1.7--4.1\,$\times$}) compared with SSD ({4.3--4.9\,$\times$} to achieve optimal performance).
Moreover, its performance is {2.3--4.1\,$\times$} higher than SSD.

%% file: body/cluster.tex
\section{Improved Cluster Index Design}
\label{sec:cluster}

\noindent
To break the trade-off between index space and performance, 
we propose two system designs for placing the cluster index efficiently on the second-tier memory: \\[-10pt]
\begin{enumerate}[leftmargin=*,leftmargin=10pt,itemindent=0pt]
    \item \textbf{Decoupled index layout (\textsection{\ref{sec:design-cluster-basic}}).}
    We store the vectors and clusters separately on the second-tier memory. 
    Thus, the index build algorithm can replicate vectors (with high factors) in clusters 
    with negligible index size amplification.    
    \\[-15pt]

    \item \textbf{Cluster-aware grouping (\textsection{\ref{sec:group}}). }
    Decoupled layout inevitably introduces irregular I/Os on the second-tier memory, 
    which can degrade performance if there are too many. 
    Therefore, we propose a cluster-aware grouping algorithm 
    to reorganize the vector layout post-index build to reduce such I/Os.    
\end{enumerate}

\subsection{Basic design: Decoupled layout}
\label{sec:design-cluster-basic}

\begin{figure}[!t]
        \hspace{-10pt}
        \includegraphics[scale=1.13]{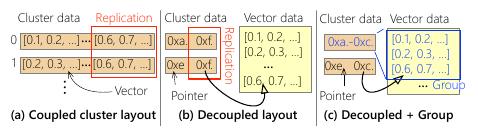}  \\[1pt]
        \begin{minipage}{1\linewidth}
        \caption{\small{%
        An overview of how we improve the layout of the vector index on the second-tier memory, 
        from the traditional coupled design (a), to a decoupled design (b), 
        and to a design with grouping (c).
        }}
        \label{fig:decoupled}
        \end{minipage} \\[-5pt]
    \end{figure} 

\noindent
To resolve the dilemma of high-performance and low cluster index size, 
we deduplicate the cluster index with a decoupled layout.
As shown in {\fig{fig:decoupled}},
 unlike the existing cluster index that directly stores the (replicated) vectors in each cluster (a), 
we store vectors separately with the cluster data, 
and only store the addresses of vectors in each cluster (b).
As a result, for replicated vectors, we only replicate the address.
Since the address size is orders of magnitude smaller than a vector (8\,B vs. 100--384\,B vectors), 
replicating addresses has trivial storage overhead to the index (see (d)--(f) in {\fig{fig:cluster-break}}).

The execution flow of the index search is the same as with the decoupled layout. 
The only difference is that when reading a cluster, 
we will first read the vector addresses belonging to the cluster, 
and then issue separate reads to fetch the vectors.
Those reads are small random reads with 100--384\,B payloads, 
which is not efficient for traditional storage like SSD but is
much more efficient on the second-tier memory. 

\stitle{Problem: too many irregular I/Os. }
Though the decoupled layout can reduce the index size,
it also decouples the original {0.1--2.2\,MB} I/O 
in to {60--68}\,$\times$ small random I/Os (100--384\,B).
More specifically, 98\% of the I/Os in the workloads are small random I/Os, 
which are also inefficient on the second-tier memory (see {\fig{fig:hybrid-access}}).
As a result, we observe a {27--39\%} performance drop on all the workloads when increasing the replication factor
(see {\fig{fig:cluster-break}} (a)--(c)).
Note that improving the replication factor is still important 
for an optimal index size. 

\input{./body/group}

\begin{figure*}[!ht]
    \hspace{-5pt}
    \includegraphics[left, scale=1.1]{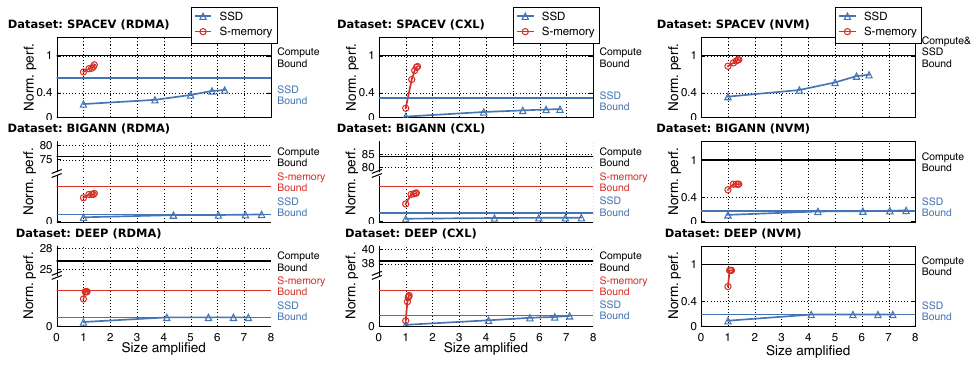} \\[1pt]
    \begin{minipage}{1\linewidth}
        \caption{\small{
        End-to-end cluster index performance. 
        S-memory states for second-tier memory used.
        We omit the S-memory bound in experiments on the SPACEV dataset and NVM 
        as it is too large (45--430\,$\times$).        
        }}
    \label{fig:cluster-scale}
\end{minipage} \\[-15pt]
\end{figure*}

\subsection{Overall Performance}
\label{sec:cluster-perf}

\vspace{-1.ex}
\stitle{Effects of grouping. }
{\fig{fig:cluster-break}} shows the effects of cluster-aware grouping. 
Specifically, grouping improves the performance on various datasets 
by {1.3--1.7}\,$\times$, 
and can achieve close or even higher performance than the replicated performance. 
The improved performance is due to the reduced number of vectors transferred: 
it reduces {62--80\%} I/O requests in the chosen workloads.

\stitle{End-to-end performance. }
{\fig{fig:cluster-scale}} shows the index performance respective to index amplification 
caused by replicating vectors
with all the above optimizations enabled\footnote{\footnotesize{Due to limited memory on our platform, 
we only report the performance of 100M datasets for CXL.}}.
Like graph, we calculate the computation bound by performing all calculations using in-memory data 
and bandwidth bound by with the peek bandwidth measured in \fig{fig:random-access}.
On second-tier memory, 
cluster index can achieve both high performance and low space utilization with our cluster-aware grouping.
Specifically, it only requires 16--22\% of the size of the SSD index,
and achieves {1.4--7.1}\,$\times$ higher throughput.
Unlike graph index, the bottleneck of the three datasets is quite different on cluster index.
SPACEV is bottlenecked by computation in all settings,
 as it requires reading about only {0.1}\,MB per query, far less than the other two datasets (0.7\,MB and 2.2\,MB).
 This is because it is a skewed dataset where vectors are close. 
Both BIGANN and DEEP are bottlenecked by second-tier memory bandwidth,
as cluster index requires reading more vectors for the search than graph index due to 
its coarse-grained index nature. 
Interestingly, on NVM, BIGANN and DEEP are still bottlenecked by the computation power of the CPU, 
because the equipped CPU is weak ({see Table~\ref{tab:hardware-info}}).

\begin{figure*}[!t]
    \hspace{-5pt}
    \includegraphics[left, scale=1.1]{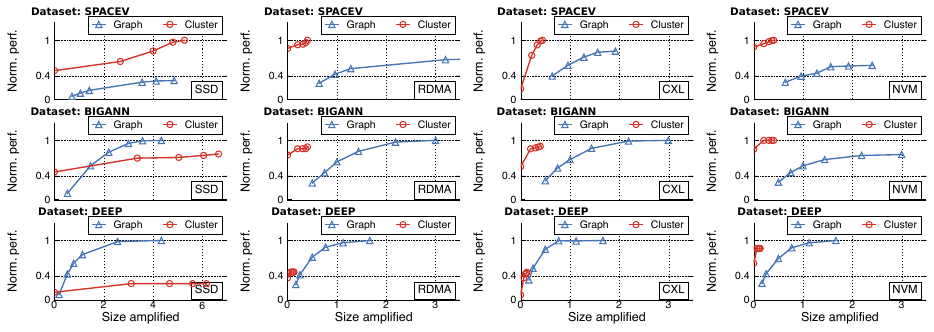} \\[1pt]
    \begin{minipage}{1\linewidth}
        \caption{\small{
        End-to-end comparison of graph and cluster index performance on (a) SSD, (b) RDMA, (c) CXL, and (d) NVM.
        }}
    \label{fig:overall}
\end{minipage} \\[-5pt]
\end{figure*}

%% file: body/group.tex
\subsection{Cluster-aware grouping}
\label{sec:group}

\input{./body/group-variables-tab.tex}

\noindent
To reduce the number of small I/Os in the workloads, 
we propose vector grouping atop of the decoupled design
to minimize the number of small random I/Os.
By grouping the vectors belonging to the same cluster together 
and storing them in adjacent storage, 
we can use one large I/O to read all vectors in the group (see {\fig{fig:decoupled}} (c)).
The group is done post the index build. 

Since we don't replicate vectors, 
there must exist clusters that we cannot use one I/O to read all its vectors,
e.g., when searching the cluster 1, we still need to use a separate small I/O to 
read the vector data belonging to address \texttt{0xc}.
Therefore, the performance of grouping heavily depends on how the vectors are grouped. 
The vector group problem aims to determine 
which vectors should be grouped together to minimize small IOs in a given workload. 
It is equivalent to the problem of assigning vectors to specific cluster groups. 
Assume we have a vector database with a set of vectors $V$ and a set of clusters $C$. 
We use $P_{i,j}$ to denote whether $i_{th}$ vector has been assigned to a group.

We employ an integer linear program (ILP) to find the best group of $P_{i,j}$. 
The objective is to minimize the IO requests sent to the MN.
Since the number of IOs depends on the access frequency of each cluster, 
we use $h_{j}$ to denote the access frequencies of clusters. 
The frequencies can be obtained by monitoring the workload in the background, 
and we can dynamically change the layout according to frequency changes. 
Since dynamic adjustment is not the focus of our work, we will omit it in this paper.

Putting it all together, we formulate the problem as follows:
\begin{equation}
    \begin{aligned}
    & \text{\emph{minimize}}   & & \sum_{j}^{|C|} h_{j} \cdot (1 + \sum_{i}^{|V|}P_{ij})  \\
    & \text{\emph{subject t}o} & & \text{\emph{Cluster constraints}} \\
    &                          & & \text{\emph{Group constraints}}
    \end{aligned}
    \end{equation}

\stitle{Cluster constraints.}
According to the cluster index design, 
a vector must be assigned to one cluster, 
while being replicated to multiple adjacent clusters of the assigned cluster.
We formulate the constraints as follows:
\[
\left\{
\begin{alignedat}{2}
&0 \leq A_{i,j} \leq 1, &\quad &i \in [0, |V|), \, j \in [0, |C|) \\
& \sum_{j}^{|C|} A_{i,j} \geq 1, &&i \in [0, |V|)
\end{alignedat}
\right.
\]

\stitle{Group constraints. }
The group algorithm must assign a vector to 
exactly one cluster's group:
\[
\left\{
\begin{alignedat}{2}
&0 \leq P_{i,j} \leq 1, &\quad &i \in [0, |V|), \, j \in [0, |C|) \\
& \sum_{j}^{|C|} P_{i,j} = 1, &&i \in [0, |V|)
\end{alignedat}
\right.
\]

\stitle{Solve the algorithm at billion-scale.}
One simple way to solve the problem is to use an off-the-shelf solve~\cite{solver}.
However, for billion-scale vectors, it takes non-trivial time to solve the problem (at least polynomial complexity). 
Observing the simple structure of the problem, 
we can use a simple greedy algorithm to find the optimal solution.
Specifically, for a vector that has been replicated to multiple clusters,
assigning it to the cluster with the highest access frequency is the optimal choice. 
This is because, 
informally, assigning it to a less frequently accessed cluster will increase the number of I/Os.
Due to space limitation, we omit the detailed proof. 
As a result, we can quickly solve the problem even for billion-scale vectors in <90 minutes.

%% file: body/group-variables-tab.tex
\begin{table}[!t]
    \vspace{2.2mm}
    \begin{minipage}{1\linewidth}
        \caption{\small{
            Variables used in the vector group algorithms. 
            The optimization variable is determined by the algorithm, 
            and constant variables come from the index build.
        }}
    \label{tab:group-variables}
    \end{minipage} \\[-1pt]
    \hspace*{-5mm}
    \centering
    \small{
    \resizebox{.94\linewidth}{!}{
    \ra{1.1}

    \begin{tabular}{l p{6cm}} 
        \toprule
        \multicolumn{2}{l}{\uline{Constant variables:}}          \\
        $A_{i,j}$ & If $A_{i,j}$ is 1,  
        it means that the $i_{th}$ vector is partitioned or replicated to the $j_{th}$ cluster's. \\
        $h_{i}$   & The access frequency of the $i_{th}$'s cluster.  \\
        \midrule 
        \multicolumn{2}{l}{\uline{Optimization variable:}}      \\
        $P_{i,j}$  & If $P_{i,j}$  is 1, it means that the $i_{th}$ vector belongs to the $j_{th}$ cluster's group. \\
        \bottomrule
        \end{tabular}

    }
    } 
    \end{table} 

%% file: body/overall.tex
\section{An end-to-end comparison of graph and cluster index on second-tier memory}
\label{sec:study}

\vspace{-1.ex}
\noindent
An interesting question faced by developers that use vector search is 
what is the right index---graph or cluster---to choose. 
For SSD-based vector indexes,
a common belief is that graph index is better in storage 
as it won't need many edges to achieve a high accuracy~\cite{DBLP:journals/corr/abs-2401-02116}. 
Therefore, developers tend to choose graph index for its smaller index size.
In comparison, cluster index requires sufficient replications to achieve the same accuracy level, 
which is also observed in the paper. 
For cluster index, a common belief is that it is better in performance as its coarse-grained 
access pattern is suitable for the SSD. 
As a result, developers tend to choose cluster index for its higher performance~\cite{SPANN,DBLP:conf/sosp/XuLLXCZLYYYCY23}.

\input{./body/compute-tab}

On second-tier memory, the findings are completely different with our improved index design: \\[-18pt]
\begin{enumerate}[leftmargin=*,leftmargin=10pt,itemindent=0pt]
    \item Graph index typically has a better IO efficiency on second-tier memory. 
    \item Cluster index has a small index footprint on second-tier memory. \\[-10pt]
\end{enumerate}

\stitle{Performance comparison. }
{\fig{fig:overall}} shows the end-to-end performance comparison of graph and cluster index on SSD, RDMA, CXL, and NVM
with respective to the index size. 
On BIGANN and DEEP datasets, 
Graph index is {2.1, 2.2, and 1.2$\times$} faster than cluster on RDMA, CXL, and NVM, respectively. 
The core reason is that graph reads less data than cluster: 
on these datasets, graph reads {95--97\%} fewer bytes than the cluster (see Table~\ref{tab:compute}). 
Though it decouples such reads into fine-grained I/Os ({512--1024\,B}), 
they are efficient on second-tier memory, so 
it is different from the common belief that graph index is slower than cluster. 
Note that without our software pipeline (\textsection{\ref{sec:graph}}), 
graph still cannot outperform cluster due to access delays when traversing the graph. 

On SPACEV, cluster index is still {1.5, 1.2, and 1.7$\times$} faster than graph on RDMA, CXL, and NVM, respectively.
This is because it is a dataset with clusters close in distance. 
As a result, it only needs to search a few clusters and thus has better performance due to the efficient grouped vector reads enabled by our grouping method.

\stitle{Space comparison. }
Thanks to our decoupled layout with group,
we nearly eliminated the space amplification caused by the replicated vectors. 
As a result, 
on SPACEV, BIGANN, and DEEP datasets, 
cluster index can achieve optimal performance with only {0.1--0.4}\,$\times$ index size amplification.
This is {56--68\%} smaller than the graph index. 
For such amplification, 
cluster index is a good candidate for index with small amplification on second-tier memory.

%% file: body/compute-tab.tex
\begin{table}[t]
    \vspace{2.2mm}
    \begin{minipage}{1\linewidth}
    \caption{\small{{
        The computational intensity and average bytes read per query vary on different datasets for vector indexes.
    }}} 
    \label{tab:compute}
    \end{minipage} \\[3pt]
    \centering
    \small{
    \resizebox{.99\linewidth}{!}{
    \ra{1.18}
\begin{tabular}{l|ccccc} \toprule
    & \multicolumn{2}{c}{\textbf{Computation intensity}}                        & \multicolumn{1}{c}{} & \multicolumn{2}{c}{\textbf{KBytes Read/query}}                                  \\ \cline{2-3} \cline{5-6} 
\textbf{Datasets} & \multicolumn{1}{c}{\textbf{Graph}} & \multicolumn{1}{c}{\textbf{Cluster}} & \multicolumn{1}{c}{} & \multicolumn{1}{c}{\textbf{Graph}} & \multicolumn{1}{c}{\textbf{Cluster}} \\ \cline{1-3} \cline{5-6} 
\textbf{SPACEV}   &    22                                &     10                                 &                     &  32                                  &      139                                \\
\textbf{BIGANN}   &    29                                &     8                                 &                      &  38                                  &      745                                \\
\textbf{DEEP}     &    19                                &     3                                 &                      &  60                                  &      2,374                                \\
\bottomrule
\end{tabular}  
    }
    }
\end{table}

%% file: body/related.tex
\section{Related Work}
\label{sec:related}

\vspace{-1.ex}
\stitle{Other vector indexes. }
A large amount of effort in indexing algorithms has been dedicated to enhancing the efficiency of vector search. 
Besides the graph and cluster indexes extensively discussed in this paper,
tree-based indexing organizes data in a hierarchical tree structure, 
such as KD-tree\cite{kd-tree}, Ball-tree\cite{ball-tree} or R-tree\cite{r-tree}.
Hash-based indexing\cite{Spectral-Hashing,DBLP:journals/pvldb/HuangFZFN15,DBLP:conf/vldb/GionisIM99} 
uses hashing functions to map similar data points to the same buckets, 
allowing for approximate nearest neighbor queries via hash table lookups.  
Unfortunately, these indexes cannot either scale to large-scale (e.g., billion-scale) datasets or 
provide sufficient accuracy~\cite{similarity-study,ADBV}.

\stitle{Vector database. }
Several systems have been developed to handle the complexities of storing, indexing, 
and searching high-dimensional vector data. 
AnalyticDB-V\cite{ADBV}, PASE\cite{PASE} and VBASE\cite{vbase} integrate vector search into traditional database, 
allowing for complex queries through SQL syntax. 
Milvus\cite{milvus} is an open-source vector database designed to handle embedding vectors converted from unstructured data. 
Faiss\cite{faiss} and Annoy\cite{annoy} developed libraries for efficient vector search. 
Pinecone\cite{pinecone} is a cloud-native vector database that simplifies 
the process of building applications with vector search capabilities.
SPFresh\cite{DBLP:conf/sosp/XuLLXCZLYYYCY23} implements LIRE to support incremental in-place updates for billion-scale vector search.
Elasticsearch\cite{Elasticsearch} and Solr\cite{Solr} can support vector search using algorithms like HNSW in distributed search engines. 
They are orthogonal to our work, and 
they use the index built by our study to further enhance the performance of vector search with small index amplifications.

\stitle{New hardware for vector search.} Recently, with the rise of new computation hardware like GPUs and FPGAs,
and new I/O hardware like NVM and CXL, many system designers have explored the potential of utilizing them to accelerate vector search~\cite{DBLP:journals/tbd/JohnsonDJ21, DBLP:conf/nsdi/ZhangL0L024, cagra, cxl-anns}.
For example, faiss library has proposed a GPU implementation of cluster index to accelerate batched queries. RUMMY is a system that implements efficient compute-IO pipeline to accelerated cluster-based
vector search. For graph-based vector search, CAGRA presents a new graph index design that has affinity for parallelism during index construction phase on GPU. They also achieve a performance 
improvement on CPU-based graph index. CXL-ANNS is a system that utilizes CXL and NDP to accelerate graph-based vector search.

%% file: concl.tex
\section{Conclusion}
\label{sec:concl}

\noindent
Existing large-scale vector indexes on SSD face a dilemma of high performance versus low index amplification, 
which we attribute to a mismatch between workload requirements and high-performance SSD access patterns.
Such a dilemma—--though challenging to address from an algorithm's perspective, 
can be systematically addressed by a co-design with emerging second-tier memory.
Our improved index design can achieve {1.5--5.6}\,$\times$ performance improvements with {15--62\%} less index amplification for graph,
and {1.4--5.9}\,$\times$ performance improvements with {92--98\%} less index amplification for cluster.
These results show that second-tier memory is a promising storage medium for vector indexes. 